# Wireless Keypads – A New Classroom Technology Using Enhanced Multiple-Choice Questions

**RAY A. BURNSTEIN**
*Illinois Institute of Technology, Chicago, IL 60616, USA*

**LEON M. LEDERMAN**
*Illinois Institute of Technology, Chicago, IL 60616*
*Illinois Mathematics and Science Academy, Aurora, IL 60506, USA*

**ABSTRACT**

*This article discusses the advantages of using wireless keypads in the lecture/classroom. This new technology requires multiple-choice (MC) questions to mate with the keypad entry features of these devices. The format of the traditional MC response is constrained to five choices and only one 'best' response is allowed. For this reason, we propose enhancements to the traditional MC question. This enhanced MC question allows as many as ten answers. The answers can vary in their degree of correctness and can be assigned partial credit. By combining wireless keys and multiple-choice questions, we can readily perform both formative and summative assessments of student learning. Examples and classroom applications are presented.*

## Introduction

The use of wireless keypads and similar devices in the lecture/classroom converts a passive audience to active learners. This is accomplished by embellishing the lecture format with numerous teacher presented questions requiring immediate responses from students via keypads. A histogram of class responses is then projected, thus providing the teacher with rapid feedback. However, the keypads require multiple-choice (MC) questions to mate with the keypad entry. The format of the traditional MC response lacks flexibility since it is conventionally constrained to a limited number of choices and only one 'correct' response is allowed. In this article we discuss keypad usage and several important issues relating to the development and value of multiple-choice questions. We review the procedures for writing the traditional/standard MC question and we propose enhancements. These new questions which we call 'enhanced MC' questions encourage as many as ten answers. The answers can vary in their degree of correctness and can be assigned partial credit. Such features add flexibility to keypad use in the classroom. In addition, keypads provide improved classroom assessment tools as well as creating a stimulating interruption to the traditional lecture format for a wide spectrum of classrooms from secondary school levels through to college/university levels.

## Features and Role of the Traditional Multiple-Choice Question

New technology has spawned devices like wireless keypads that have changed, and will further change the classroom and teaching procedures. Wireless keypad hardware mates well with multiple-choice questioning. MC questions can be used in classroom quizzes as well as major examinations. Multiple-choice testing has assumed this important role because it has proven to be an efficient and effective tool for assessing students' learning over large numbers of students, many disciplines and varying levels of knowledge. Consequently, one finds that major national and professional achievement tests such as the SAT, ACT, GRE and MCAT[1] examinations are constructed using the multiple-choice format. This format also has a potentially huge future role in the testing associated with the (United States) No Child Left Behind Act.[2] A major virtue of multiple choice tests is that they can be given frequently, require little time to grade, and thereby provide a continuous stream of assessments during a semester. In addition to assessing learning, multiple-choice questions help generate classroom discussion and can also be used to obtain information on student opinions and attitudes. Criticisms of multiple-choice testing correctly point out the increased difficulty in probing a student's ability to formulate opinions and organize critical arguments. These skills are better assessed by, e.g. essays, portfolios, non-algorithmic problem tests and research or project performance.

Given the important role that multiple-choice testing occupies, it is not surprising that there are many publications that describe the traditional methods for the construction of multiple-choice questions.[3] In this article we reference previous work and propose alternative methods for constructing multiple-choice questions with improved features and discuss their use with the technology of classroom response systems. We believe this information will be valuable to the many teachers who wish to construct multiple-choice questions with little background knowledge of the traditional procedures.

## Format of a Traditional Multiple-Choice Test Item

The existence of a traditional/standard multiple choice format is evidenced by the fact that the tens of millions of multiple choice tests produced each year appear to have common elements and format. A traditional multiple-choice test item as illustrated in Table 1, consists of two basic parts: (1) a problem (*stem*), and (2) a list of suggested solutions (*alternatives/options*). The stem may be in the form of a question or a problem. The suggested solutions (alternatives/options) ordinarily contain one correct or best alternative (*answer*) and a number of incorrect alternatives (*distractors*).

Distractors[4] should appear as plausible solutions to the problem for those students who have not achieved the objective being measured by the test item. Conversely, the distractors should appear to be implausible solutions for those students who have achieved the objective. Only the best alternative (the answer) should be plausible to these knowledgeable students. Table 1 illustrates the construction of a traditional multiple-choice question with a, b, c, and d, the distracters for the question. Numerous examples of multiple-choice questions using physics concept questions can be found in reference.[5]

Table 1. Structure of a traditional multiple-choice question.[5]

Two metal balls are the same size but one weighs twice as much as the other. The two balls are dropped from the roof of a one story building at the same instant of time. The time it takes the the balls to reach the ground is: — **Stem**

Distracter a. about half as long for the heavier ball as for the lighter ball
Distracter b. about half as long for the lighter ball as for the heavier ball
Answer       about the same for both balls
Distracter c. considerably less for the heavier ball not necessarily half as long
Distracter d. considerably less for the lighter ball not necessarily half as long

— **Alternatives Options**

There is a detailed article in the literature about how to write the stems and how to write the distracters.[4] These suggested procedures appear well defined; for example the prescription that distractors are best constructed when comparable in length, complexity, and grammatical form to the answer, as seen in Table 1.

### The Classification of Multiple Choice Questions

Assessment efforts are directed to determining whether students have a grasp of important concepts. There are basically two types of assessment procedures, formative and summative. In the case of formative assessment, this information provides continuous feedback to the instructor and then to the students. In the case of summative assessment, the information is used to determine the extent of learning. The universe of assessment techniques is very large, but for the most part, tests of student progress are usually restricted to multiple-choice, problem solving, or essay questions. It is useful to classify the range/level of understanding covered by these testing questions. One way to assess foundational thinking skills is by using Bloom's Taxonomy.[6] This taxonomy has evolved into a classic work that organizes the levels of learning into six categories ranging from simple to complex. The levels can be classified as: Knowledge, Comprehension, Application, Analysis, Synthesis, and Evaluation.

There are many representations of Bloom's hierarchy and a condensed version follows.

The use of Bloom's Taxonomy allows one to classify the cognitive level of the multiple-choice questions presented. We later use this classification scheme to categorize our sample questions.

**Table 2.** The classification of Bloom's cognitive levels.

| Level | Characteristic Student Behaviors |
| --- | --- |
| 1. Knowledge | Remembering; memorizing; recognizing |
| 2. Comprehension | Interpreting; describing in one's own words |
| 3. Application | Problem-solving; applying information to produce a result |
| 4. Analysis | Subdividing to show how something is put together; identifying motives |
| 5. Synthesis | Creating a unique, original product |
| 6. Evaluation | Making value decisions about issues; resolving controversies |

**Advantages and Limitations of Multiple-Choice Questions**

Some of the advantages of traditional multiple-choice questions have already been described in this article. The limitations of traditional multiple-choice questions are subtle but nonetheless significant. There is limited flexibility in preparing and grading multiple-choice questions. For example, the common limitation to four or five choices is not essential and probably reflects the fact that the grading vehicle, usually an optical scanner[7] only handled a limited number of choices. In addition, the scanning machines only allowed credit for one correct answer. In contrast, recent keypads systems and other real-time data entry systems are computer based and the software used for grading allows more possibilities.[8-10] For example, up to ten alternatives are possible for each question with most keypad systems. Furthermore, the software allows each different answer to have a different score which could be related to its 'correctness'. Also new questions can be created on-the-fly to take advantage of the rapid feedback provided by a keypad response system. It is a huge advance for the teacher to be able to quiz and receive feedback from hundreds of students in seconds. All these factors suggest that at this time it is feasible and advantageous to modify and optimize the traditional multiple choice question.

**Enhanced multiple-choice questions**

Variations in the traditional multiple-choice question format have appeared.[11-15] We label this type of question as; an enhanced or partially correct multiple-choice (EMC) question.

The main advantages and characteristics of these enhanced EMC questions are:

- EMC removes the restriction of limited choices per question
- EMC removes the limitation of one 'correct' answer per question
- EMC introduces the possibility of awarding partial credit for answers
- EMC with more choices allows one to simulate short essay type responses
- EMC makes it easier to create higher cognitive level questions (Bloom's Taxonomy)

Our EMC questions are physics based since they were developed for use with wireless keypads in introductory college physics courses at Illinois Institute of Technology.[8] However, the basic style of the questions is applicable to almost every science course and with many types of response systems. Our experience is that while some EMC questions do take more time to compose and answer, we find that students benefit from the associated discussion and welcome the partial credit features associated with these questions. Brief examples follow.

Problem #1 is used to assess whether students understand the underlying physics of the ballistic pendulum problem (a bullet fired into a block that is constrained as the bob of a pendulum). One answer is requested, but partial credit is given for some of the alternatives. In this case, a multiple-choice problem in the traditional format could not accomplish the quizzing goal as well as the EMC question because of the limited number of choices available. There are also elements of an essay type question here and this quasi-essay question can be often used to focus on basic principles. Most important, is that the choices presented probe the students' understanding of a key concepts such the conservation laws of energy and momentum. When the teacher presents this question he/she gets immediate feedback on the student' understanding of concept and can immediately alter ongoing lecture if appropriate.

1 A bullet of 10gm strikes a ballistic pendulum (a bullet fired into a block that is constrained as the bob of a pendulum). The block has a mass of 2kg.

After the collision, the bullet emerges from the block with negligible velocity and the center of mass of the pendulum rises a distance of 12cm. If you were asked to calculate the bullet's initial speed, what are the physics conservation laws which you would use to produce a solution to the problem?

a. Use conservation of energy for the collision of the bullet and block
b. Use conservation of momentum for the collision of the bullet and block
c. Use conservation of energy for the motion of the block
d. Use conservation of momentum for the motion of the block
e. Use conservation of momentum for the collision of the bullet and block and conservation of momentum for the motion of the block
f. Use conservation of momentum for the collision of the bullet and block and conservation of energy for the motion of the block
g. Use conservation of energy for the collision of the bullet and block and conservation of momentum for the motion of the block
h. Use conservation of energy for the collision of the bullet and block and conservation energy for the motion of the block
i. Not enough information given to produce a solution

($f$ = correct, b, c = partially correct)
This EMC question represents Bloom's Taxonomy level 3 – application/solve.

Problem #2 is an EMC question best suited for discussion use. One answer is requested; find the word that fits least well in the list of words:

a. Wood
b. Oil
c. Coal
d. Grass
e. Energy
f. Banana
g. Chair
h. Tree

(e = correct, energy is an invented concept; all other words are objects that one can see or touch.

This question represents Bloom's Taxonomy level 5 – synthesis (Bloom's cognitive level)

### The Use of Multiple Choice Questions in Large Classes/Lectures with Keypad Systems

Traditional multiple-choice questioning, as has been discussed, is one of the most commonly used procedures for classroom assessment.[16] This assessment usually takes the form of summative assessment because its purpose is to give a summary of achievement at various times and can be performed in large college classes because of the automated grading features provided by optical scanning devices. Wireless keypad systems and other polling devices[10] can replace the optical scanners since the questions can be similar and the grading is also automated. In addition, wireless keypads can also be used for formative assessment. The essential feature of formative assessment is that it inserts rapid feedback into the teaching and learning process. For this reason, keypads and other on-line devices are uniquely well suited for formative assessment in large classes because there is built-in prompt feedback measured in seconds which is otherwise not possible for a large audience. Quizzing with keypads[8] involves interrupting a 60 to 90 minute lecture 5-15 times with multiple-choice quiz questions and evaluating and recording the results. The results are transmitted from the computer and projected to the class. The teacher can immediately respond in a number of ways present a new or related question, initiate a group discussion etc. When the questions are complicated or when the questions are presented involving group discussion (peer learning mode), fewer questions can be asked during a class session.[13] On the other hand, more questions can be presented when the questions are simple, e.g. to test whether the reading assignment was done.

Formative assessment produces significant and often substantial learning gains according to the research of Black & William.[17] Another approach to understanding learning gains has been pioneered by Hake.[18] He has concluded, based on substantial experimental data, that learning gains are correlated to the extent of teacher-student interaction referred to as 'interactive engagement'. Actually, the use of formative assessment and 'interactive engagement' in the classroom are closely related activities since both activities involve classroom feedback. A careful reading of Hake[19] and Black and Wiliam[17] reveals this association.

In our classes we have often used formative assessment in a general way by asking the following keypad question (in an anonymous mode) at appropriate times during a lecture to instantly get student feedback on the progress of a lecture topic.

QUESTION: Evaluate the physics topic just presented
 a. Score 7-10- needs no further clarification
 b. Score 5-7- a few points need clarification
 c. Score 3-5- many points need clarification
 d. Score 1-3- can't even ask a question because I understand so little

Our feedback/data from this type of question indicates that students are alert and often request clarifications. In response, the teacher can quickly suggest additional questions, more clarifications, etc. This type of opinion question is not included in the classification scheme of Bloom's Taxonomy and might be just regarded as a student feedback/opinion question. None-the-less, this type of question is useful for the formative assessment process. This question is usually presented in the anonymous mode; where the teacher can tabulate the responses but does not know the identity of the responder.

In our classes many of the questions asked are presented in a non-anonymous mode, that is, the answer is graded and recorded in a computer file for each student. These questions are graded such that no answer receives 0 points, any answer receives 3 points; the correct answer receives 10 points; and when there is a partially correct answer, that answer would receive variable credit. With this particular grading system a student is encouraged to record an answer, and that answer would be expected to be his/her best choice. The accumulated scores for each student over an entire semester (summative assessment) can be used at the teacher's discretion to count for a portion of the course grade. Even in the extreme case where the grade is determined principally by a 'high-stakes test', for example, as is in the case of the No-Child-Left-Behind yearly examination, the accumulated scores from semester-long keypad testing can be a useful supplement.[20] Keypad data is useful in classes, especially large classes, to give early information about weak, poorly prepared or non-attending students.

## Summary and Conclusions

The traditional multiple-choice question has been the predominant format for the vast majority of achievement test questions. This standard format involves well-defined question generation procedures with which some teachers are unfamiliar. For this reason a study of the traditional procedures, as summarized in this paper, is a useful first-step in the process of test preparation. In this article, we also illustrate and identify different types of multiple-choice questions (enhanced multiple-choice/EMC). These types of questions are produced by altering the traditional multiple-choice format in a number of different ways. We maintain that these EMC questions are especially useful now because the software available with current computer driven classroom response systems allows a wider variety of questions and grading possibilities.

The ease and ability of obtaining rapid feedback from real-time student quizzing indicates that keypads and other on-line devices can play an important role in providing formative assessment data.[17] Further, it is our experience that summative assessment data can be obtained at the same time. As a result, keypads and multiple-choice questions can provide versatile and practical assessment procedures that can be applied to a wide spectrum of science and even non-science classrooms from secondary schools through to college/university levels.

## Acknowledgements

We would like to thank Professors Richard Hake and Eugenia Etkina for helpful comments. The National Science Foundation provided partial support for initial wireless keypad development.

## References

1. SAT, ACT, GRE, MCAT are acronyms for, respectively, Scholastic Assessment Test, American College Test, Graduate Record Exam, and Medical College Admissions Test; details can be obtained at the respective websites.
2. NCLB U.S. Dept. of Education, Public Law PL 107-110, the No Child Left Behind Act of 2001, <http://www.ed.gov/policy/elsec/leg/esea02/index.html
3. J. Carneson, G. Delpierre, and K. Masters, Designing and managing multiple choice

questions, University of Cape Town, 2006 (unpublished), <http://web.uct.ac.za/projects/cbe/mcqman/mcqman01.html>
4. R.B. Frary, *Practical Assessment, Research & Evaluation* 4(11) (1995) and references therein, <http://pareonline.net/getvn.asp?v=4&n=11>
5. This example is from D. Hestenes, D., M. Wells and G. Swackhammer, *Phys. Teach.* 30, 141(1992)
6. B.S. Bloom, Taxonomy of Educational Objectives, Handbook I: The Cognitive Domain, (New York: David McKay Co. Inc., 1956)
7. e.g. Scantron Corp. Irvine, CA, online at http://www.scantron.com
8. R A. Burnstein and L. M. Lederman, *Phys. Teach.* 39, 8 (2001), an updated unpublished list of wireless keypad systems is available from R. A. Burmstein
9. R.R. Hake, A comprehensive compilation of references for Classroom Communication Systems, 2004 (unpublished), http://listserv.nd.edu/cgi-bin/wa?A2 = ind0412&L=pod&O=D&P=24855.
10. Banks ed. "*Audience response systems in higher education: Applications and cases*, (Hershey, PA: Information Science Publishing, 2006). This book contains a current and comprehensive review of the subject
11. R.A. Burnstein and L. M. Lederman, *AAPT Announcer* 30(4) 85 (2000), Examples of our enhanced/partially correct multiple-choice questions (EMC) were first detailed in this presentation
12. R.J. Dufresne and W. J. Gerace, *Phys. Teach.* 42, 109 (2004)
13. E. Mazur, *Peer Instruction, A user's manual,* 1997 (Upper Saddle River, NJ: Prentice Hall, 1997)
14. T.L. O'Kuma, D. P. Maloney, and C. J. Hieggelke, *Ranking Task Exercises in Physics,* (Upper Saddle River, NJ, Prentice Hall, 1999)
15. E.F. Redish, Teaching *Physics: with the Physics Suite,* (New York: John Wiley & Sons, Inc., 2003)
16. T.A. Angelo, and K. P. Cross, *Classroom assessment techniques: A Handbook for College Teachers,* 2nd ed., (San Francisco, Jossey-Bass, 1993)
17. P. Black, and D. Wiliam, *Phi Delta Kappan* 80(2), 139 (1998)
18. R.R Hake, *Am. J. Phys.* 66(1) 64 (1998), <http://www.physics.indiana.edu/~sdi/ajpv3i.pdf>, see ref. 19 for a crucial companion paper
19. R.R. Hake, Interactive-engagement methods in introductory mechanics courses 1998 (unpublished), http://www.physics.indiana.edu/~sdi/IEM-2b.pdf
20. L.M. Lederman and R A. Burnstein, *Phi Delta Kappan,* 87(6), 429 (2006)